\documentclass[final,1p,times]{elsarticle} 
\usepackage{graphicx} 
\usepackage{amssymb} 
\usepackage{amsthm} 

\usepackage{epsf,epsfig}
\usepackage{sidecap}

\journal{Nuclear Physics A} 
\begin{document} 

\begin{frontmatter} 


\title{Direct Photon-Hadron Correlations Measured with PHENIX}

\author{Megan Connors for the PHENIX Collaboration}

\address{Department of Physics and Astronomy, Stony Brook University, Stony Brook NY, 
11794}

\begin{abstract} 

Direct photon-hadron correlations greatly improve our ability to
perform jet tomography in heavy-ion collisions because the momentum
of the direct photon can be used to constrain the initial momentum
of the opposing jet. By comparing the spectrum of away-side hadrons
observed in heavy ion collisions to the spectrum seen in nucleon
collisions we can quantify the medium modification to the
fragmentation function due to energy loss of the away-side parton.


High $p_{T}$ direct photon-hadron correlations have been measured with
the PHENIX detector using a statistical subtraction method to remove
the photon contribution from meson decays. The
increased integrated luminosity in the most recent Au+Au RHIC run at
$\sqrt{s_{NN}}$ = 200 GeV provides substantially improved
statistical precision and enhances the kinematic reach. 
These measurements are compared to PHENIX p+p results
and several theoretical models of energy loss. In addition, we
compare direct photon-hadron and $\pi^{0}$-hadron correlations.

\end{abstract} 

\end{frontmatter} 



Direct photon - jet correlations are considered a ``golden channel''
for studying jet tomography since the transverse momentum of the jet is
approximately balanced by the transverse momentum of the photon and photons do
not interact with the medium created at RHIC. The
lack of medium modification to the direct photons has previously
been shown by the PHENIX direct photon $R_{AA}$ measurement~\cite{raa}, which shows 
that $R_{AA}$ of direct photons is consistent with 1 out to at least 14 GeV/c in 
$p_T$. Direct photon - hadron correlations also allow measurement of the 
fragmentation function of the opposing parton~\cite{wang}. The fragmentation 
function is $D_{q}(z)=dN(z)/(N_{evt}dz)$, where $z = p_{hadron}/p_{jet}$. For $\gamma_{direct}-h$ correlations, $z$ is 
approximately known, since the transverse momentum of the photon has the same magnitude as the 
opposing jet momentum to leading order. At NLO, however, the $k_{T}$ effect is important~\cite{justin}.


The measurement of these correlations, however, is complicated by the need to 
separate correlations involving direct photons from those involving inclusive photons. In heavy 
ion collisions, one of the largest sources of photons is decay
photons, mostly from $\pi^{0}\rightarrow\gamma\gamma$. Direct photons are defined as 
all
photons which do not result from a hadronic decay process since the decay
photons are the only contribution subtracted from the inclusive sample.
Prompt photons are the direct result of a hard scattering such as Compton
scattering ($qg\rightarrow q\gamma$) or quark-antiquark annihilation. Since Compton 
scattering is dominant, the away-side fragmentation is mostly quark jet 
fragmentation.
Photons from jet fragmentation and medium-induced photons also
contribute to the inclusive sample. To measure $\gamma_{direct}-h$ correlations, the 
contribution of
decay photons are subtracted from inclusive $\gamma-h$ correlations
via a statistical subtraction method. More details on the method and results from the 
2004 data have been presented elsewhere~\cite{ppg090}. First, inclusive
$\gamma-h$ correlations are determined by measuring the angle
between photons detected in the electromagnetic calorimeter and
charged hadrons measured in the PHENIX tracking system. The measured inclusive photon 
yield can be written as:

\begin{equation} \label{eqn:yinclusive}
  Y_\mathrm{inc} = 
\frac{N_\mathrm{dir}Y_\mathrm{dir}+N_\mathrm{dec}Y_\mathrm{dec}}{N_\mathrm{inc}}.
\end{equation}

The conditional yield here is defined as $Y=N_{pair}/N_{trig}$. To determine the 
decay contribution, first, $\pi^{0}-h$ correlations are
measured. To translate $\pi^{0}-h$ correlations to $\gamma_{decay}-h$, a Monte Carlo 
simulation is used to determine the probability that a pion decays into a photon in a 
particular $p_{T}$ bin.  A correction is applied in $Au+Au$ to account for the 
contribution from the $\eta$ decays based on the $p+p$ measurement. Higher mass 
decays are accounted for in the systematic error based on PYTHIA. Once the yield from 
$\gamma_{decay}-h$ is determined, the subtraction described in Equation 
\ref{eqn:subtraction}, which is Equation \ref{eqn:yinclusive} rearranged, is 
performed, where $R_{\gamma}=N_\mathrm{inc}/N_\mathrm{dec}$ and has been previously 
measured~\cite{raa}.

\begin{equation} \label{eqn:subtraction}
  Y_\mathrm{dir} = \frac{R_{\gamma}Y_\mathrm{inc}-Y_\mathrm{dec}}{{R_\gamma}-1}.
\end{equation}

\begin{figure}[ht]
    \begin{center}
      \epsfig{file=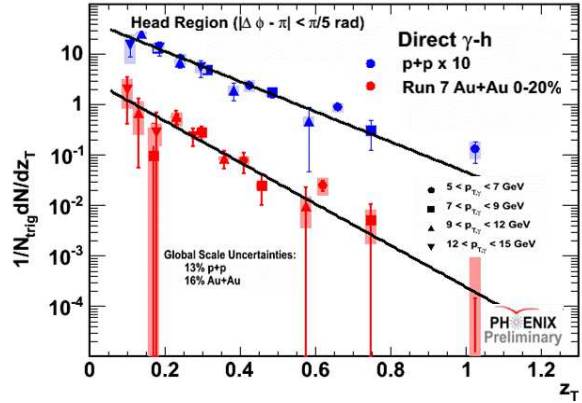,scale=0.35}
      \caption{{\rm (Color online) The $z_{T}$ distribution for $p+p$ scaled by a factor of 10 (blue) and $Au+Au$ (red).}}
      \label{dndzt}
    \end{center}
  \end{figure}

To measure the modification of the fragmentation function of the opposing parton, the 
away-side yield of the resulting $\gamma_{direct}-h$ jet function is measured in the 
head region, $|\Delta\phi-\pi|<\pi/5$, and plotted as a function of $z_{T}$. 
For correlations between trigger and associated particles, $z_{T} = 
\langle p_{T,assoc}\rangle/\langle p_{T,trig}\rangle$. For $\gamma_{direct}-h$, $z_{T} = 
\langle p_{T,h}\rangle/\langle p_{T,\gamma}\rangle \cong \langle p_{T,h}\rangle/\langle p_{T,jet}\rangle$ and is used to approximate $z$. 
The $z_{T}$ distributions are shown in Figure \ref{dndzt}. The upper set of points are for $p+p$ collisions from the 2005 and 
2006 data combined and the lower set are for $Au+Au$ collisions from the 2007 data. The 2007 data 
set is a factor of four larger than the previously shown 2004 data. The improved data set allows for a measurement at higher $p_{T,h}$, extending the results to 
higher $z_{T}$.

All $p_{T}$ bins appear to obey approximate $z_{T}$ scaling. Therefore, all points for each collisional system were fit with the function, 
$\frac{dN}{dz_{T}}=Ne^{-bz_{T}}$. For $p+p$ the slope was measured to be 
$b=6.89\pm0.64$. The slope in $Au+Au$ is $b=9.49\pm1.37$ which exceeds that in 
p+p, as expected for modified fragmentation in the QCD medium due to constant 
fractional energy loss of the away-side parton. However, the statistical 
uncertainties in these data limit the significance of the difference to only 1.3 $\sigma$.

\begin{figure}[ht]
    \begin{center}
      \epsfig{file=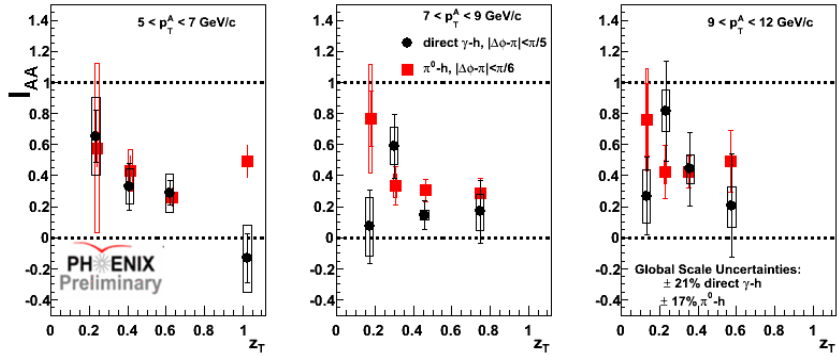,scale=0.4}
      \caption{{\rm The $I_{AA}$ measured in the head regions for $\gamma_{dir}-h$ and $\pi^{0}-h$ as a function of $z_{T}$ for three different $p_{T,trig}$.}}
      \label{pi0iaa}
    \end{center}
  \end{figure}

To quantify the suppression observed in the $\gamma_{dir}-h$ channel, the $I_{AA}$ is 
measured. The $I_{AA}=Y_{Au+Au}/Y_{p+p}$ is the ratio of the conditional yield observed in 
$Au+Au$ to that in $p+p$. As with $R_{AA}$, $I_{AA}=1$ would indicate no 
modification. Figure \ref{pi0iaa} shows suppression for $Au+Au$ yield since 
$I_{AA}<1$. This suppression is compared to the suppression measured in $\pi^{0}-h$ 
in the same figure. With the possible exception of the lowest and highest $z_{T}$ 
points, the $I_{AA}$ of $\gamma_{dir}-h$ and $\pi^{0}-h$ are remarkably consistent. 
This is surprising since $\pi^{0}-h$ should include more gluon jets which are more 
suppressed. However, $\pi^{0}-h$ suffers a surface bias, which reduces the 
suppression observed, while $\gamma_{dir}-h$ does not suffer such bias since the 
$\gamma_{dir}$ is not suppressed by the medium.

\begin{figure}[ht]
    \begin{center}
      \epsfig{file=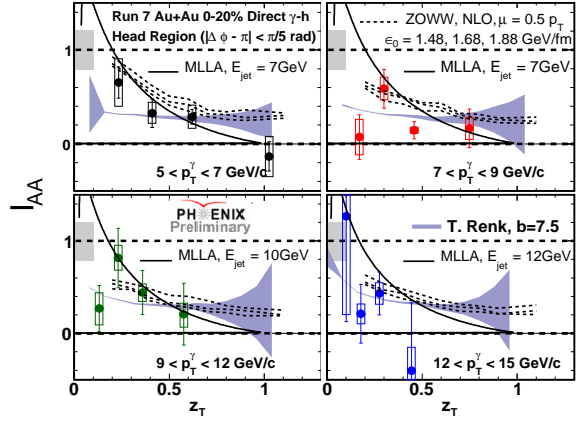,scale=0.42}
      \caption{{\rm The $I_{AA}$ measured in the head regions for $\gamma_{dir}-h$ 
(circles) compared to three different predictions~\cite{zoww, renk, borghini}.}}
      \label{theoryiaa}
    \end{center}
  \end{figure}

The $I_{AA}$ for $\gamma_{dir}-h$ is also compared to several of the current 
theories as shown in Figure~\ref{theoryiaa}. The ZOWW curve~\cite{zoww} shows 
suppression at high $z_{T}$ and less suppression at low $z_{T}$. This shape results 
from the geometric dependence of the hard scatterings. High $z_{T}$ is more surface 
biased since it requires the associated hadron to have higher momentum while low 
momentum hadrons (low $z_{T}$ pairs) probe deeper into the medium, reducing the 
observed suppression. Renk's curve~\cite{renk}, however, does not have a strong 
$z_{T}$ dependence because he allows for fluctuations in his energy loss model, 
washing out the geometric dependence effects. Finally, the data are compared to MLLA 
results from~\cite{borghini} in which the energy loss goes into production of low 
momentum particles. This enhances the low $z_{T}$ region such that at very low 
$z_{T}$, the $I_{AA}$ is above unity. The current measurements can not fully rule out 
any of these theories. However, these measurements will benefit from additional 
statistics from future runs and possibly improvements of the subtraction method. 

\begin{figure}[htb]
    \begin{center}
      \epsfig{file=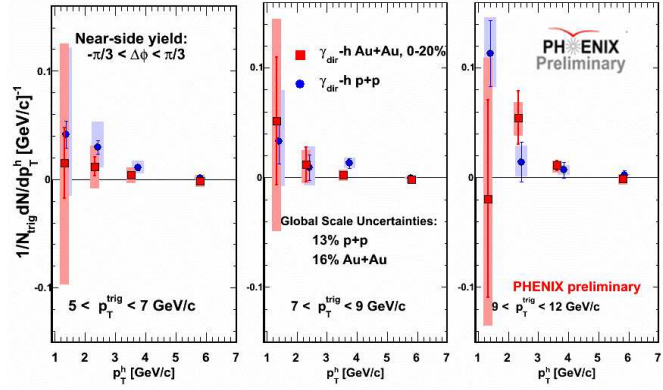,scale=0.42}
      \caption{{\rm The nearside yield in $p+p$ (circles) and $Au+Au$ (squares) as a function of $p_{T,h}$ for different $p_{T,\gamma}$.}}
      \label{nearside}
    \end{center}
  \end{figure}

We also study $\gamma_{dir}-h$ correlations at small opening angles. The near-side 
yield of associated particles for $p+p$ and $Au+Au$ are shown in Figure~\ref{nearside}. All points are consistent with little to no yield on the near-side; 
it may be non-zero since fragmentation and possible medium induced photons 
contribute. The $Au+Au$ data also show no indication of enhancement compared to the $p+p$ points.
 
In conclusion, suppression has been observed on the away-side of the $\gamma_{dir}-h$ 
channel in $Au+Au$ collisions at 200 GeV. This suppression is remarkably consistent 
with the suppression measured in the $\pi^{0}-h$ correlations. The $z_{T}$ 
distributions shown suggest a steeper slope in $Au+Au$, $9.49\pm1.37$, compared to 
the slope, $6.89\pm0.64$, measured in $p+p$ collisions. Preliminary measurements of 
the nearside show no evidence of ehancement in $Au+Au$ compared to $p+p$. With 
increased statistics and extended kinematic reach, experimental results of 
$\gamma_{dir}-h$ are moving toward precision measurements of energy loss.



\end{document}